\documentclass
[preprint,byrevtex,pre,nofootinbib,10pt,balancelastpage,titlepage,bibnotes,twocolumn]{revtex4}%
\usepackage{amsfonts}
\usepackage{amsmath}
\usepackage{amssymb}
\usepackage{graphicx}%
\providecommand{\U}[1]{\protect\rule{.1in}{.1in}}

\ifx\pdfoutput\relax\let\pdfoutput=\undefined\fi
\newcount\msipdfoutput
\ifx\pdfoutput\undefined\else
\ifcase\pdfoutput\else
\ifx\paperwidth\undefined\else
\ifdim\paperheight=0pt\relax\else\pdfpageheight\paperheight\fi
\ifdim\paperwidth=0pt\relax\else\pdfpagewidth\paperwidth\fi
\fi\fi\fi
\begin{document}
\preprint{UATP/1804}
\title{Jensen Inequality and the Second Law}
\author{P.D. Gujrati}
\email{pdg@uakron.edu}
\affiliation{Department of Physics, Department of Polymer Science, The University of Akron,
Akron, OH 44325}

\begin{abstract}
Jensen's Inequality (JIEQ) has proved to be a major tool to prove the
consistency of various fluctuation theorems with the second law in microscopic
thermodynamics. We show that the situation is far from clear and the reliance
on the JIE may be quite misleading in general.

\end{abstract}
\date{August 20, 2018}
\maketitle

1. \textsc{Introduction }The Jensen inequality (JIEQ) \cite{Cover} has become
popularized \ recently in modern nonequilibrium (NEQ) thermodynamics as a tool
to imply that various (integral) fluctuation theorems (FT)
\cite{Searles,Mansour,Seifert,Harris} of the form%
\begin{equation}
\left\langle e^{-\Delta\Phi}\right\rangle _{\text{TE}}%
=1,\label{IntegralFluctuationTheorem}%
\end{equation}
such as the Jarzynski identity, Crooks theorem, Seifert's entropy generation
theorem, etc. are consistent with the second law. These FTs are determined by
the trajectories $\left\{  \gamma_{k}\right\}  $ and their probabilities
$\left\{  p_{\gamma_{k}}\right\}  $; see various reviews
\cite{Esposito,Campisi,Seifert,Jarzynski-Rev}. The collection $\left\{
\gamma_{k}\right\}  $ forms the \emph{trajectory ensemble} (TE) and defines
the average $\left\langle \bullet\right\rangle _{\text{TE}}$. Jensen's
inequality, a purely \emph{mathematical} result to obtain the inequality
$\left\langle \Delta\Phi\right\rangle _{\text{TE}}\geq0$ from Eq.
(\ref{IntegralFluctuationTheorem}), is extensively used to argue for the
conformity of the FTs with the second law, a \emph{macroscopic} result in
physics. It thus follows that the only use of the JIEQ is to justify that the
FTs can describe NEQ processes so as to make them quite suitable to gain
insight into the second law. As the choice of $\left\{  p_{\gamma}\right\}  $
is not unique, see below, the relationship of $\left\langle \Delta
\Phi\right\rangle _{\text{TE}}$\ with its \emph{thermodynamic average}, to be
denoted simply by $\left\langle \Delta\Phi\right\rangle $, is not clear, since
the second law must only refer to thermodynamic averages such as $\left\langle
\Delta S_{0}\right\rangle $ for the entropy change of an isolated system
$\Sigma_{0}$ for which the second law results in the inequality $\left\langle
\Delta S_{0}\right\rangle \geq0$. Another consequence of the second law is the
dissipated work in an isothermal process%
\begin{equation}
\Delta R_{\text{diss}}\doteq\left\langle \Delta R\right\rangle -\Delta
F=T_{0}\Delta_{\text{i}}S\geq0,\label{DissipatedWork}%
\end{equation}
where $\left\langle \Delta R\right\rangle $ is the thermodynamic average work
(see Eq. (\ref{Thermodynamic Average}) for a proper definition) done on the
system \cite{Note1,Landau} during a process, $\Delta F$ is the free energy
change, and $\Delta R_{\text{diss}}$ is the \emph{dissipated work
}\cite{Note-R}. As an example of a FT, Jarzynski \cite{Jarzynski} derived
\begin{equation}
\left\langle e^{-\beta_{0}\Delta R_{\text{diss}}}\right\rangle _{\text{0}%
}\doteq%
{\textstyle\sum\limits_{k}}
p_{k0}e^{-\beta_{0}\Delta R_{k\text{diss}}}=1,\label{JarzynskiEquality}%
\end{equation}
where $\Delta R_{k\text{diss}}\doteq\Delta R_{\gamma_{k}}-\Delta F,\Delta
R_{\gamma_{k}}$ denotes the work done on the system $\Sigma$ along the
trajectory $\gamma_{k}$, the suffix $0$ refers to a special averaging with
respect to the initial equilibrium (EQ) microstate probabilities $\left\{
p_{k0}\right\}  $ replacing $\left\{  p_{\gamma_{k}}\right\}  $, and $k$
refers to the initial microstate of $\gamma_{k}$. The above identity is known
as the \emph{Jarzynski identity }(JE). We will refer to $\left\langle
\bullet\right\rangle _{0}$ as the \emph{Jarzynski average} in this work.

It is implicitly assumed in the current literature that $\left\langle
\Delta\Phi\right\rangle _{\text{TE}}\equiv\left\langle \Delta\Phi\right\rangle
$. What is the significance of $\left\langle \Delta\Phi\right\rangle
_{\text{TE}}\geq0$ if $\left\langle \Delta\Phi\right\rangle $ does not refer
to a quantity that must obey the second law (see the discussion later for such
a quantity)? Indeed, we will establish here that using the JIEQ can be
misleading in suggesting that the FT applies to NEQ processes or that
$\left\langle \Delta\Phi\right\rangle _{\text{TE}}$ satisfies the second law,
while in fact they do not. To the best of our knowledge, this issue has not
been discussed in the literature despite the wide use of the JIEQ. We
establish that (i) the \emph{trajectory ensemble average} (TEA) $\left\langle
\bullet\right\rangle _{\text{TE}}$ may or may not be the same as the
\emph{thermodynamic average} $\left\langle \bullet\right\rangle $, and (ii)
even when the two are the same, $\left\langle \Delta\Phi\right\rangle $ may
have nothing to do with the second law. As a consequence, the consequence of
the JIEQ need not refer to the second law, thus casting doubts on its utility
for FTs.

For simplicity, we consider a "work-process" $\mathcal{P}_{0}$\ on a system
$\Sigma$ as proposed by Jarzynski \cite{Jarzynski}. \ It is an arbitrary
process$\ $over $(0,\tau_{\text{eq}})$ between two EQ macrostates \textsf{A
}and \textsf{B} at the same inverse temperature $\beta_{0}=1/T_{0}$; here,
$\tau_{\text{eq}}$ is the time needed to reach the EQ macrostate \textsf{B}.
The system is driven (the \emph{driving stage}$\ \mathcal{P}$) over
$(0,\tau),\tau\leq\tau_{\text{eq}}$, by $\widetilde{\Sigma}_{\text{w}}$ and is
then allowed to equilibrate (the \emph{reequilibration stage}$\ \overline
{\mathcal{P}}$) due to interaction with $\widetilde{\Sigma}_{\text{h}}$ only
over $(\tau,\tau_{\text{eq}})$. For simplicity, we assume that during
$\mathcal{P}$, $\Sigma$ is not in thermal contact with $\widetilde{\Sigma
}_{\text{h}}$. We denote by $\widetilde{\Sigma}$ the combination
$\widetilde{\Sigma}_{\text{h}}\cup\widetilde{\Sigma}_{\text{w}}$ and the
combination $\Sigma\cup\widetilde{\Sigma}$ by $\Sigma_{0}$, which is an
isolated system. All quantities pertaining to $\Sigma$\ have no suffix, and
those pertaining to $\widetilde{\Sigma}$ ($\Sigma_{0}$) with a tilde (suffix
$0$). For concreteness, we assume the work process to change the volume $V(t)$
of the system by applying an external pressure $P_{0}$, but the arguments are
valid for any external "work" process. The system-intrisic (SI) \cite{Note-SI}
pressure for the $k$th microstate $\mathfrak{m}_{k}$ will be denoted by
$P_{k}\doteq-\partial E_{k}/\partial V$, where $E_{k}$ is the microstate
energy, an SI-quantity. The difference $\Delta P_{k}\doteq P_{k}-P_{0}$
denotes the ubiquitous force imbalance (FI) between the external and induced
internal forces that is normally nonzero even in equilibrium
\cite{Gujrati-GeneralizedWork,Gujrati-GeneralizedWork-Expanded}. Therefore, to
discard FI in an irreversible process is \emph{counter-productive}. We find it
convenient to use Prigogine's modern notation, which is highly suitable in NEQ
thermodynamics
\cite{deGroot,Prigogine,Gujrati-II,Gujrati-Entropy2,Gujrati-Stat}.

2. \textsc{Jensen}$^{^{\prime}}$\textsc{s Inequality }Consider a \emph{convex}
function $\Phi(X)$ of a random variable $X$, and let \textsf{E} be an
expectation operator such as $\left\langle \bullet\right\rangle _{0}%
,\left\langle \bullet\right\rangle _{\text{TE}}$, etc. Then, the inequality%
\begin{equation}
\mathsf{E}(\Phi(X))\geq\Phi(\mathsf{E}(X)) \label{JensenInequality}%
\end{equation}
is known as Jensen's inequality (JIEQ) for $\Phi(X)$. For the JE,
$\left\langle \bullet\right\rangle _{0}$ represents \textsf{E} so the JIEQ
results in $\left\langle \Delta R\right\rangle _{0}-\Delta F\geq0$. By
exploiting an ad-hoc assumption $\left\langle \Delta R\right\rangle
_{0}=\left\langle \Delta R\right\rangle $ without offering any justification,
Jarzynski \cite{Jarzynski} has argued that the JE results in $\left\langle
\Delta R\right\rangle \geq\Delta F$ in accordance with the second law; see Eq.
(\ref{DissipatedWork}). The use of the JIEQ\ has become widespread to
establish consistency with the second law by exploiting a similar ad-hoc
assumption $\left\langle \Delta R\right\rangle _{\text{TE}}=\left\langle
\Delta R\right\rangle $ such as by Crooks \cite{Crooks}, Seifert
\cite{Seifert,Seifert-PRL,Seifert-EPJ,Jarzynski-EPJ} and many others. The
argument is \emph{crucial} since it indirectly "justifies" the results to be
nonequilibrium results. The assumption $\left\langle \Delta R\right\rangle
_{\text{TE}}=\left\langle \Delta R\right\rangle $ is never ever explicitly
mentioned but seems to have been accepted by all workers without ever been justified.

3. \textsc{Thermodynamic Ensemble Averages }In general, an EQ or NEQ
\emph{ensemble average} (EA) is defined instantaneously, and\ requires
identifying (a) the elements (microstates $\left\{  \mathfrak{m}_{k}\right\}
$) of the ensemble and (b) their instantaneous probabilities $\left\{
p_{k}\right\}  $. The average is \emph{uniquely} defined over $\left\{
\mathfrak{m}_{k}\right\}  $ using $\left\{  p_{k}\right\}  $\ at each instant,
which we identify as the \emph{instantaneous ensemble average} (IEA). Let
$O_{k}$ be some extensive quantity pertaining to $\mathfrak{m}_{k}$. The
instantaneous thermodynamic average $\left\langle O\right\rangle $ is defined
\cite{Prigogine,Landau} as
\begin{equation}
\left\langle O(t)\right\rangle \doteq%
{\textstyle\sum\nolimits_{k}}
O_{k}(t)p_{k}(t). \label{Thermodynamic Average-General}%
\end{equation}
We will usually not show the time $t$ unless clarity is needed. In
thermodynamics, it is common to simply use $O$ for the average but it may
cause confusion in some cases. We will use macroquantity for the average $O$
and microquantity for $O_{k}$. The average energy $E\equiv\left\langle
E\right\rangle $ is such an average \emph{system-intrinsic} (SI)
macroquantity. The infinitesimal thermodynamic work $dR\equiv\left\langle
dR\right\rangle $ done on the system and the work done by the system
$dW\equiv\left\langle dW\right\rangle $ represent such an average
instantaneous macroquantities; the former is \emph{medium-intrinsic} (MI)
quantity and the latter a SI quantity. The first law during $dt$ is expressed
as a sum of two \emph{system-intrinsic} (SI) contributions%
\begin{equation}
d\left\langle E\right\rangle =%
{\textstyle\sum\nolimits_{k}}
E_{k}dp_{k}+%
{\textstyle\sum\nolimits_{k}}
p_{k}dE_{k}. \label{FirstLaw}%
\end{equation}
The first sum represents the \emph{generalized heat} $dQ=TdS$ while the second
sum represents $-dW$, the \emph{generalized work}
\cite{Gujrati-GeneralizedWork,Gujrati-Stat,Gujrati-GeneralizedWork-Expanded}
in terms of the SI microwork $dW_{k}=-dE_{k}$ done by $\mathfrak{m}_{k}$.
These generalized macroquantities should not be confused with the exchanged
macroheat and macrowork $d_{\text{e}}Q$ and $d_{\text{e}}W$, respectively.
Their differences are $d_{\text{i}}Q$ and $d_{\text{i}}W$, respectively, with
an important identity of their magnitudes $d_{\text{i}}Q=d_{\text{i}}W\geq0$.
We also observe that during generalized work, $\left\{  E_{k}\right\}  $
change but not $\left\{  p_{k}\right\}  $; during generalized heat, $\left\{
p_{k}\right\}  $ change but not $\left\{  E_{k}\right\}  $. This allows us to
treat work and heat separately.

4. \textsc{Trajectory Ensemble Averages }The uniqueness inherent in Eq.
(\ref{Thermodynamic Average-General}) may not hold for the TEA $\left\langle
\bullet\right\rangle _{\text{TE}}$, which we now discuss. Let $\gamma_{k}%
$\ denote the trajectory followed by $\mathfrak{m}_{k}$ during its evolution
along $\mathcal{P}_{0}$. The average cumulative change $\left\langle \Delta
O_{\mathcal{P}_{0}}\right\rangle $ along a process $\mathcal{P}_{0}$, we
suppress the suffix TE for simplicity, is obtained by \emph{integrating}
$\left\langle dO(t)\right\rangle $ over the process between $t^{\prime}=0$ and
$t^{\prime}=\tau_{\text{eq}}$:
\begin{equation}
\left\langle \Delta O_{\mathcal{P}_{0}}(t)\right\rangle \doteq%
{\textstyle\int\nolimits_{\mathcal{P}_{0}}}
\left\langle dO\right\rangle \doteq%
{\textstyle\sum\nolimits_{k}}
{\textstyle\int\nolimits_{\gamma_{k}}}
p_{k}(t^{\prime})dO_{k}(t^{\prime}); \label{Thermodynamic Average}%
\end{equation}
we will use $\Delta O_{\mathcal{P}_{0}}(\tau_{\text{eq}})$ or simply $\Delta
O_{\mathcal{P}_{0}}$ or $\Delta O$ for $\left\langle \Delta O_{\mathcal{P}%
_{0}}\right\rangle $ unless clarity is needed. We note that $\mathfrak{m}_{k}$
retains its identity during its evolution along $\mathcal{P}_{0}$ as indicated
by the sum; no transition between different microstates is allowed. We can
also introduce the cumulative change $\Delta O_{\gamma_{k}}(t)$ $\doteq%
{\textstyle\int\nolimits_{\gamma_{k}}}
dO_{k}$ along $\gamma_{k}$ over the interval $\left(  0,t\right)  $, and
rewrite the above equation as%
\begin{equation}
\Delta O\doteq%
{\textstyle\sum\nolimits_{\gamma_{k}}}
p_{\gamma_{k}}^{(\text{O})}(\tau_{\text{eq}})\Delta O_{\gamma_{k}}%
(\tau_{\text{eq}}), \label{Thermodynamic process average}%
\end{equation}
where we have introduced the \emph{trajectory probability} $p_{\gamma_{k}%
}=p_{\gamma_{k}}^{(\text{O})}(\tau_{\text{eq}})$ in terms of $\Delta
O_{\gamma_{k}}(\tau_{\text{eq}})$:
\begin{equation}
p_{\gamma_{k}}^{(\text{O})}(\tau_{\text{eq}})\doteq%
{\textstyle\int\nolimits_{\gamma_{k}}}
p_{k}(t)dO_{k}(t)/\Delta O_{\gamma_{k}}(\tau_{\text{eq}})=%
{\textstyle\int\nolimits_{\gamma_{k}}}
p_{k}(t)dx_{k}^{(\text{O})}(t); \label{TrajectoryProbability-General}%
\end{equation}
here $x_{\gamma_{k}}^{(\text{O})}(t)\doteq\Delta O_{\gamma_{k}}(t)/\Delta
O_{\gamma_{k}}(\tau_{\text{eq}}),t\leq\tau_{\text{eq}}$. We note from Eq.
(\ref{Thermodynamic process average}) that $\Delta O$ can also be treated as
the thermodynamic average with respect to the trajectory probability set
$\{p_{\gamma_{k}}^{(\text{O})}(\tau)\}$. Using $dR_{k}(t)$ and $dW_{k}(t)$ for
$dO_{k}(t)$, we obtain the average accumulated work $\Delta R$ done on and
$\Delta W$ by the system, respectively, in terms of the respective trajectory
probabilities: \
\begin{subequations}
\begin{align}
\Delta R(\tau)  &  \doteq%
{\textstyle\sum\nolimits_{\gamma_{k}}}
p_{\gamma_{k}}^{(\text{R})}(\tau_{\text{eq}})\Delta R_{\gamma_{k}}%
(\tau),\label{ThermodynamicAverageWork-R}\\
\Delta W(\tau_{\text{eq}})  &  \doteq%
{\textstyle\sum\nolimits_{\gamma_{k}}}
p_{\gamma_{k}}^{(\text{W})}(\tau_{\text{eq}})\Delta W_{\gamma_{k}}%
(\tau_{\text{eq}}); \label{ThermodynamicAverageWork-W}%
\end{align}
the probabilities are determined by $x_{k}^{(\text{R})}(t)\doteq\Delta
R_{\gamma_{k}}(t)/R_{\gamma_{k}}(\tau),t\leq\tau,x_{k}^{(\text{R})}%
(t)\doteq1,\tau\leq t\leq\tau_{\text{eq}}$, and $x_{k}^{(\text{W})}%
(t)\doteq\Delta W_{\gamma_{k}}(t)/\Delta W_{\gamma_{k}}(\tau_{\text{eq}%
}),t\leq\tau_{\text{eq}}$, respectively, in Eq.
(\ref{TrajectoryProbability-General}); here, we have used the fact that
$dR_{k}(t)$ is nonzero only during the driving stage $(0,\tau\leq
\tau_{\text{eq}})$. One can similarly define a trajectory probability
$p_{\gamma_{k}}^{(\text{I})}(\tau_{\text{eq}})$ by using $x_{k}^{(\text{t}%
)}(t)\doteq t/\tau_{\text{eq}}$ over $\gamma_{k}$:%
\end{subequations}
\[
p_{\gamma_{k}}^{(\text{Av})}(\tau_{\text{eq}})\doteq%
{\textstyle\int\nolimits_{\gamma_{k}}}
p_{k}(t)dx_{k}^{(\text{t})}(t).
\]
This average (over time) probability is determined by $\gamma_{k}$ alone and
can be identified as the \emph{intrinsic }trajectory probability. We observe
that $\Delta R(\tau)\equiv\Delta R(\tau_{\text{eq}})$. It should be evident
that the three probabilities are not the same. In other words, there is no
unique trajectory probability $p_{\gamma_{k}}$ as said earlier.

The trajectory $\gamma_{k}$ is determined by a single microstate
$\mathfrak{m}_{k}$, and proves useful in the thermodynamic macroworks $\Delta
W$ or $\Delta R$. We can also consider a \emph{mixed} trajectory (mT)
$\overline{\gamma}_{k\rightarrow m}(t)$ as a sequence $\left\{  \mathfrak{m}%
_{j}\right\}  _{j=0,1,\cdots,n}$ of microstates starting at $\mathfrak{m}%
_{0}=k$ at $t_{0}=0$ and terminating at $\mathfrak{m}_{n}=m$ at time
$t=t_{n}=\tau_{\text{eq}}$; the microstate $\mathfrak{m}_{l},l<n,$ appears at
time $t=t_{l}<t_{l+1}\leq\tau_{\text{eq}}$. Consider the time interval
$\delta_{l}\doteq(t_{l},t_{l+1})$, which we divide into an earlier interval
$\delta_{l}^{\prime}\doteq(t_{l},t_{l}^{\prime}<t_{l+1})$ and a later interval
$\delta_{l}^{\prime\prime}\doteq(t_{l}^{\prime},t_{l+1})$. During $\delta
_{l}^{\prime}$, $\mathfrak{m}_{l}$ does not change as microwork $\delta
R_{\mathfrak{m}_{l}}$ is performed by $\widetilde{\Sigma}_{\text{w}}$; no
microheat is transferred. During $\delta_{l}^{\prime\prime}$, no microwork
$\delta R_{\mathfrak{m}_{l}}$ is performed by $\widetilde{\Sigma}_{\text{w}}$
but microheat is transferred, which changes $\mathfrak{m}_{l}$ to
$\mathfrak{m}_{l+1}$. For $\mathfrak{m}_{l}=k,\forall l$, $\overline{\gamma
}_{k\rightarrow k}(\tau)$ reduces to $\gamma_{k}$. The probability
$p_{\overline{\gamma}_{k\rightarrow m}}$ is given by
\[
p_{\overline{\gamma}_{k\rightarrow m}}=p_{k0}T(\left\{  \mathfrak{m}%
_{l}\right\}  _{l=1,2\cdots,n}\mid k)
\]
in terms of the multistate conditional probability $T(\left\{  \mathfrak{m}%
_{l}\right\}  _{l=1,2\cdots,n}\mid k)$ and the \emph{initial} probability
$p_{k0}$. The corresponding $\left\langle \bullet\right\rangle _{\text{TE}}$
with respect to $p_{\overline{\gamma}_{k\rightarrow m}}$ is obtained by
replacing $\gamma_{k}$ and $p_{k}$ by $\overline{\gamma}_{k\rightarrow m}$ and
$p_{\overline{\gamma}_{k\rightarrow m}}$, respectively,\ in Eq.
(\ref{Thermodynamic Average}) and summing over all $\overline{\gamma
}_{k\rightarrow m}$. It is very common to assume that the sequence $\left\{
\mathfrak{m}_{l}\right\}  _{l=0,1,\cdots,n}$ forms a (memoryless) Markov (M)
chain so that $T(\left\{  \mathfrak{m}_{l}\right\}  _{l=1,2\cdots,n}\mid k)$
can be expressed as a product of two-state transition probabilities $T_{ij}$
to determine the Markov approximate $p_{\overline{\gamma}_{k\rightarrow m}%
}^{(\text{M})}$. Using $p_{\overline{\gamma}_{k\rightarrow m}}^{(\text{M})}$,
the Markov average external work over $t\in(0,\tau_{\text{eq}})$ is%
\begin{equation}%
\begin{tabular}
[c]{l}%
$\Delta R^{(\text{M})}(\tau_{\text{eq}})\doteq%
{\textstyle\sum\nolimits_{\left\{  \mathfrak{m}_{l}\right\}  }}
p_{\overline{\gamma}_{k\rightarrow m}}(\tau_{\text{eq}})\Delta R_{\overline
{\gamma}_{k\rightarrow m}}(\tau_{\text{eq}})$\\
$\ \ \ \ \ \ \ \ \ =%
{\textstyle\sum\nolimits_{l<n}}
{\textstyle\sum\nolimits_{\mathfrak{m}_{l}}}
p_{\mathfrak{m}_{l}}\delta R_{\mathfrak{m}_{l}}(\delta_{l}^{\prime}),$%
\end{tabular}
\ \label{Markov-Chain-Approx}%
\end{equation}
where $\Delta R_{\overline{\gamma}_{k\rightarrow n}}(\tau_{\text{eq}})\doteq%
{\textstyle\sum\nolimits_{l<n}}
\delta R_{\mathfrak{m}_{l}}(\delta_{l}^{\prime})$ and $\delta R_{\mathfrak{m}%
_{l}}(\delta_{l}^{\prime})$ is the external work done on $\mathfrak{m}_{l}$.
Thus, in the Markov chain approximation, $\Delta R^{(\text{M})}(\tau
_{\text{eq}})$ gives a \emph{discrete} approximation of the macrowork $\Delta
R(\tau_{\text{eq}})\equiv\left\langle \Delta R_{\boldsymbol{\gamma}%
}\right\rangle $ in Eq. (\ref{ThermodynamicAverageWork-R}) for which we
require $\delta_{l}^{\prime}$ to be extremely short. Otherwise, $\Delta
R^{(\text{M})}(\tau_{\text{eq}})$ and $\Delta R(\tau_{\text{eq}})$\ are very
different as we have stated earlier. For a non-Markovian process, $T(\left\{
\mathfrak{m}_{l}\right\}  _{l=1,2\cdots,n}\mid k)$ cannot be expressed as a
product of two-state transition probabilities and we must resort to the
generalization noted above of Eq. (\ref{Thermodynamic process average}).

\textsc{The Jarzynski Equality }As our first example of a TEA $\left\langle
\bullet\right\rangle _{\text{TE}}$ different from $\left\langle \bullet
\right\rangle $ in a FT, we consider the one proposed by Jarzynski
\cite{Jarzynski} noted above. Jarzynski uses the external microwork $\Delta
R_{k}$ done on $\mathfrak{m}_{k}$\ during$\ \mathcal{P}$ to prove the JE in
Eq. (\ref{JarzynskiEquality}). Here, $\Delta F=F_{\mathsf{B}}(\beta
_{0})-F_{\mathsf{A}}(\beta_{0})$, and $\Delta R_{k}\neq0$ only during the
driving stage $\mathcal{P}$; $\Delta R_{k}=0$ over $\overline{\mathcal{P}}$.
If the system at $t=\tau$ is out of equilibrium, we denote it by \textsf{b}.
The interaction with $\widetilde{\Sigma}_{\text{h}}$ during $\overline
{\mathcal{P}}$\ is to ensure that \textsf{b} turns into \textsf{B}.

The use of the JIEQ with $\left\langle \bullet\right\rangle _{0}$ for
\textsf{E} in Eq. (\ref{JarzynskiEquality}) immediately results in
$\left\langle \Delta R_{\boldsymbol{\gamma}}\right\rangle _{0}\geq\Delta F$.
Jarzynski assumes that $\left\langle \Delta R_{\boldsymbol{\gamma}%
}\right\rangle _{0}=\left\langle \Delta R_{\boldsymbol{\gamma}}\right\rangle $
and argues that the JE represents a NEQ result so that $\left\langle \Delta
R_{\boldsymbol{\gamma}}\right\rangle _{0\text{eq}}=\Delta F$ for a reversible
process and $\left\langle \Delta R_{\boldsymbol{\gamma}}\right\rangle
_{0}>\Delta F$ for an irreversible process.

We now consider a reversible process between \textsf{A }and \textsf{B,} for
which the thermodynamic macrowork $\Delta R\equiv\left\langle \Delta
R_{\boldsymbol{\gamma}}\right\rangle $ is the reversible macrowork $\Delta
R_{\text{rev}}=\Delta F$, and demonstrate by a simple example that
$\left\langle \Delta R_{\boldsymbol{\gamma}}\right\rangle $\ is not the same
as $\left\langle \Delta R_{\boldsymbol{\gamma}}\right\rangle _{0}$,
\begin{equation}
\left\langle \Delta R_{\boldsymbol{\gamma}}\right\rangle _{0}=%
{\textstyle\sum\nolimits_{k}}
p_{k0}\Delta R_{\gamma_{k}}(\tau_{\text{eq}}), \label{JarzynskiAverage0}%
\end{equation}
the Jarzynski average, except when $p_{k0}\equiv$ $p_{\gamma_{k}}^{(\text{R}%
)}(\tau),\forall\tau$.

For the calculation, we consider an ideal gas in a $1$-dimensional box of
length $L$, which expands quasistatically from $L_{\mathsf{A}}$ to
$L_{\mathsf{B}}$; we let $x\doteq L_{\mathsf{A}}/L_{\mathsf{B}}$ between
\textsf{A }and \textsf{B}. As there are no interparticle interactions, we can
treat each particle by itself. The microstates in the exclusive approach are
those of a particle in the box with energies determined by an integer
$k:E_{k}=\alpha(k/L)^{2},\alpha=\pi^{2}\hslash^{2}/2m$. Let $\beta_{0}$ denote
the inverse temperature of the heat bath. The gas remains in equilibrium at
all times and $R_{\text{diss}}=0$. The partition function at any $x$ is given
by
\[
Z(\beta_{0},L)=%
{\textstyle\sum\nolimits_{n}}
\exp(-\beta_{0}\alpha(n/L)^{2}\approx\sqrt{L^{2}\pi/4\alpha\beta_{0}}%
\]
for any $L\in\lbrack L_{\mathsf{A}},L_{\mathsf{B}}]$; in the last equation, we
have made the standard integral approximation for the sum. We then have%
\[
\beta_{0}F=-(1/2)\ln(L^{2}\pi/4\alpha\beta_{0}),E=1/2\beta_{0}.
\]

We can now compute the two work averages with $\Delta R_{k}=E_{k}%
(L_{\mathsf{B}})-E_{k}(L_{\mathsf{A}})$. For the Jarzynski average, we have%
\begin{equation}
\left\langle \Delta R_{\boldsymbol{\gamma}}\right\rangle _{0}=%
{\textstyle\sum\nolimits_{k}}
p_{k0}[E_{k}(L_{\mathsf{B}})-E_{k}(L_{\mathsf{A}})]=(x^{2}-1)/2\beta_{0},
\label{JarzynskiWork}%
\end{equation}
where we have used $E_{k}(L_{\mathsf{B}})-E_{k}(L_{\mathsf{A}})=(x^{2}%
-1)E_{k}(L_{\mathsf{A}})$. For the thermodynamic average, we use
$dE_{k}=-2E_{k}dL/L$ in Eq. (\ref{Thermodynamic Average}) to obtain%
\begin{equation}
\Delta R\equiv\left\langle \Delta R_{\boldsymbol{\gamma}}\right\rangle
=1//\beta_{0}\ln x=\Delta F. \label{ThermodynamicWork}%
\end{equation}

It should be clear that it is the thermodynamic average work $\left\langle
\Delta R_{\boldsymbol{\gamma}}\right\rangle $ that satisfies the condition of
EQ and not $\left\langle \Delta R_{\boldsymbol{\gamma}}\right\rangle _{0}$,
which is evidently different from $\left\langle \Delta R_{\boldsymbol{\gamma}%
}\right\rangle $. This, thus, contradicts the conventional assumption
$\left\langle \Delta R_{\boldsymbol{\gamma}}\right\rangle _{0}=\left\langle
\Delta R_{\boldsymbol{\gamma}}\right\rangle $. We evaluate the difference
$\left\langle \Delta R_{\boldsymbol{\gamma}}\right\rangle _{0}-\Delta F$.
Introducing $y=1-x^{2}\geq0$ for expansion, we have%
\[
\left\langle \Delta R_{\boldsymbol{\gamma}}\right\rangle _{0}-\Delta
F=[\ln(1-y)-y]//2\beta_{0}>0.
\]
The Jensen inequality is \emph{satisfied} as expected, but the above
\emph{nonnegative difference} $\left\langle \Delta R_{\boldsymbol{\gamma}%
}\right\rangle _{0}-\Delta F$ makes no statement about any dissipation in the
system, which is most certainly absent. Thus, the JIEQ makes no statement
about the second law and casts doubts on the usefulness of the indiscriminate
application of the JIEQ in FTs.

We now consider two more FTs, where the JIEQ has been used to justify
consistency with the second law.

\textsc{Crooks}$^{^{\prime}}$\textsc{Approach} Crooks \cite{Crooks}\ assumes
the evolution along $\overline{\gamma}_{k\rightarrow m}$ as a Markov process
satisfying the principle of detailed balance and divides $(0,\tau_{\text{eq}%
})$ into $n$ intervals $\delta_{l}=\delta_{l}^{\prime}\cup\delta_{l}%
^{\prime\prime},l=0,1,2,\cdots,n-1$ as described above. Microwork is performed
during $\delta_{l}^{\prime}$ and microheat is exchanged during $\delta
_{l}^{\prime\prime}$. We will not follow Crooks' derivation of the JE, which
we have carried out elsewhere \cite{Gujrati-Crooks}, but follow the
consequences of the detailed balance here. The transition probability matrix
$\mathbf{T}^{(l)}$ in $\delta_{l}$\ takes a very simple form under detailed
balance, which we denote by $\overleftrightarrow{\mathbf{T}}^{(l)}$. From the
Fundamental Limit Theorem or Doeblin's theorem\emph{ }about Markov chains
\cite{Strook,Gujtrati-Crooks}, we know that such a transition matrix is
uniquely determined with its matrix elements corresponding to $i\rightarrow j$
given by%
\begin{equation}
\overleftrightarrow{T}_{i,j}^{(l)}=p_{j\text{eq }}(V_{l+1}),\forall l<n,
\label{TransitionMatrix-Unique}%
\end{equation}
where $p_{j\text{eq}}(V_{l+1})$ is the EQ probability at fixed $V_{l+1}$ of
the $j$th microstate at time $t_{l+1}$ and ensures that the end-microstate
$\mathfrak{m}_{l+1}$ at the end of $\delta_{l}$ belongs to an EQ macrostate.
Thus, at the end of $\delta_{0}$, the EQ-microstate $\mathfrak{m}_{0}$ turns
into an EQ-microstate $\mathfrak{m}_{1}$ but the microwork done on
$\mathfrak{m}_{0}$\ is precisely $\delta R_{\mathfrak{m}_{0}}(\delta
_{0}^{\prime})$. By induction, we have a sequence of EQ-microstates
$\mathfrak{m}_{1},\mathfrak{m}_{2},\cdots,\mathfrak{m}_{n}$ and microworks
$\delta R_{\mathfrak{m}_{1}}(\delta_{1}^{\prime}),\delta R_{\mathfrak{m}_{2}%
}(\delta_{2}^{\prime}),\cdots,\delta R_{\mathfrak{m}_{n-1}}(\delta
_{n-1}^{\prime})$ and the total microwork is given by $\Delta R_{\overline
{\gamma}_{k\rightarrow n}}(\tau_{\text{eq}})$ used in Eq.
(\ref{Markov-Chain-Approx}). Using the above transition matrix
$\overleftrightarrow{\mathbf{T}}^{(l)}$, we can easily evaluate $\overline
{\gamma}_{k\rightarrow m}$:
\[
p_{\overline{\gamma}_{k\rightarrow m}}=p_{\mathfrak{m}_{0}\text{eq}}%
{\textstyle\prod\nolimits_{l=1}^{n}}
p_{\mathfrak{m}_{l}\text{eq}}%
\]
so that
\[
\left\langle e^{-\beta_{0}\Delta R}\right\rangle _{\overline{\gamma}%
}^{\text{(M)}}=%
{\textstyle\prod\nolimits_{l<n}}
\left(
{\textstyle\sum\nolimits_{\mathfrak{m}_{l}}}
p_{\mathfrak{m}_{l}\text{eq}}e^{-\beta_{0}\delta R_{\mathfrak{m}_{l}}%
(\delta_{l}^{\prime})}\right)  .
\]
We see that the Crooks process during\ $\delta_{l}$\ is no different from the
Jarzynski process and the quantity within the parentheses denotes a Jarzynski
averaging of the exponential microwork distribution over the probabilities of
the initial EQ-microstates $\{\mathfrak{m}_{l}\}$ in the interval $\delta
_{l},l=0,1,2,\cdots,n-1$. In other words, the Crooks process is a sequence of
$n$ non-overlapping mini-Jarzynski processes $\left\{  \delta\mathcal{P}%
_{0,l}\right\}  $, each over $\delta_{l},l=0,1,2,\cdots,n-1$. For each
mini-Jarzynski process, we have
\[
\left\langle e^{-\beta_{0}\delta R}\right\rangle _{l}^{\text{(M)}}%
=e^{-\beta_{0}\delta F_{l}},
\]
where the suffix $l$ denotes averaging over the initial microstate
probabilities $\{p_{\mathfrak{m}_{l}\text{eq}}\}$ and $\delta F_{l}$\ is the
change over\ $\delta_{l}$\ between EQ-macrostates. It is obvious now that
$\left\langle \delta R\right\rangle _{l}^{\text{(M)}}$ is not the same as the
thermodynamic average $\left\langle \delta R\right\rangle $, just as it was
for the Jarzynski process, unless $\delta_{l}$ is extremely small. This means
that the application of the JIEQ on $\left\langle e^{-\beta_{0}\delta
R}\right\rangle _{l}$ does not give an inequality involving thermodynamic
averages so no connection with the second law is possible.

\textsc{Seifert}$^{^{\prime}}$\textsc{s Approach}: Here, we will continue to
use a discrete formulation for simplicity. According to Seifert \cite{Seifert}%
, $S_{k}=-\ln p_{k}$ denotes the microscopic entropy and its thermodynamic
average $\left\langle S\right\rangle =%
{\textstyle\sum\nolimits_{k}}
p_{k}S_{k}$ gives the (average) entropy, commonly written as $S$. Seifert
defines the average change $dS^{(\text{S})}$ (S indicating Seifert) in terms
of $dS_{k}$,
\begin{equation}
dS^{(\text{S})}\doteq\left\langle dS\right\rangle \label{Seifert-ds}%
\end{equation}
and conjectures that $dS^{(\text{S})}$ is nothing but $d\left\langle
S\right\rangle =dS$. One can also determine $\Delta S_{k}$ as the integral of
$dS_{k}$ along the trajectory $\overline{\gamma}_{k\rightarrow m}$ and
introduce $\Delta S^{(\text{S})}\doteq\left\langle \Delta S\right\rangle
_{\overline{\gamma}}$. Similarly, we also have $dS_{0}^{(\text{S}%
)}=\left\langle dS_{0}\right\rangle $ and $\Delta S_{0}^{(\text{S})}%
\doteq\left\langle \Delta S_{0}\right\rangle _{\overline{\gamma}_{0}}$ for the
isolated system $\Sigma_{0}$ and where $\overline{\gamma}_{0}$ denotes its set
of trajectories. Seifert then derives the following equality $\left\langle
e^{-\Delta S_{0}}\right\rangle _{\overline{\gamma}_{0}}=1$. The use of the
JIEQ then results in $\Delta S_{0}^{(\text{S})}=\left\langle \Delta
S_{0}\right\rangle _{\overline{\gamma}_{0}}\geq0$, which is interpreted using
the above conjecture that $\Delta S_{0}^{(\text{S})}$ denotes $\Delta
\left\langle S_{0}\right\rangle =\Delta S_{0}=\Delta_{\text{i}}S$. Using this
interpretation, the inequality is considered a statement of the second law by
taking $\Delta S_{0}^{(\text{S})}$ to mean $\Delta_{\text{i}}S=\beta_{0}\Delta
R_{\text{diss}}$, see Eq. (\ref{DissipatedWork}). With this, Seifert provides
another proof of the JE in terms of the mixed trajectory average
\[
\left\langle e^{-\Delta S_{0}^{(\text{S})}}\right\rangle _{\overline{\gamma
}_{0}}\overset{?}{=}\left\langle e^{-\beta_{0}\Delta R_{\text{diss}}%
}\right\rangle _{\overline{\gamma}_{0}}=1.
\]
Since%
\begin{equation}
dS_{0}^{(\text{S})}\doteq\left\langle dS_{0}\right\rangle =-\left\langle
dp_{0}/p_{0}\right\rangle =-%
{\textstyle\sum\nolimits_{k}}
dp_{0k}\equiv0, \label{Gujrati-ds}%
\end{equation}
which is simply a statement of the conservation of probability, we conclude
that $dS_{0}^{(\text{S})}=0$. To determine $\Delta S_{0}^{(\text{S}%
)}=\left\langle \Delta S_{0}\right\rangle _{\overline{\gamma}}$, we follow
Eqs. (\ref{Thermodynamic Average})-(\ref{TrajectoryProbability-General}).
Since $\left\langle \Delta S_{0}\right\rangle _{\overline{\gamma}}$ is
integral of $dS_{0}^{(\text{S})}=0$, it is clear that $\left\langle \Delta
S_{0}\right\rangle _{\overline{\gamma}}=0$. Thus, the above JIEQ conclusion
$\Delta S_{0}^{(\text{S})}=\left\langle \Delta S_{0}\right\rangle
_{\overline{\gamma}_{0}}\geq0$ does not prove that it encodes the second law.
The second law requires considering the differentials $dS,d\widetilde{S}$ and
$dS_{0}$. Recalling that $dS=d\left\langle S\right\rangle =%
{\textstyle\sum\nolimits_{k}}
p_{k}dS_{k}+%
{\textstyle\sum\nolimits_{k}}
S_{k}dp_{k}$, compare with Eq. (\ref{FirstLaw}), and $dS_{k}=-dp_{k}/p_{k}$,
we have%
\begin{equation}
dS=\left\langle dS\right\rangle -\left\langle SdS\right\rangle =-\left\langle
SdS\right\rangle . \label{Gujrati-dS}%
\end{equation}
Thus, $dS^{(\text{S})}$ is not the entropy differential $dS$. Unfortunately,
this point has been overlooked.

\textsc{Conclusions }In summary, we have shown that the application of the
Jensen inequality does not at all make any statement about the second law. It
should be pointed out that while there is a consequence of the second law for
$\Delta R_{\text{diss}}$, there is no second law statement about
$dS^{(\text{S})}=0$. Thus, while in the former case, the use of the JIEQ may
provide a statement of the second law, its applications to $\left\langle
e^{-\Delta S_{0}^{(\text{S})}}\right\rangle _{\overline{\gamma}_{0}}$ has no
relationship to the second law. The conclusion is that care must be exercised
to draw any conclusion about the second law by applying the JIEQ in general, a
point that does not seem to have been appreciated.


\begin{thebibliography}{99}                                                                                               %


\bibitem {Cover}T.A. Cover and J.A. Thomas, \textit{Elements of Information
Theory}, Second Edition, John Wiley \& Sons, Hoboken, N.J. (2006).

\bibitem {Harris}R.J. Harris and G.M. Sch\"{u}tz, J. Stat. Mech. P07020 (2007).

\bibitem {Searles}E.M. Sevick, R. Prabhakar, S.R. Williams, and D. J. Searles,
Ann. Rev. Phys. Chem. \textbf{59}, 603 (2008).

\bibitem {Seifert}U. Seifert, Eur. Phys. J. B 64, 423 (2008); Rep. Prog. Phys.
\textbf{75}, 126001 (2012).

\bibitem {Mansour}M. Malek Mansour and F. Baras, Chaos, \textbf{27}, 104609 (2017).

\bibitem {Esposito}M. Esposito, U. Harbola, S. Mukamel, P. Talkner, Rev. Mod.
Phys. \textbf{81}, 1665 (2009).

\bibitem {Campisi}M. Campisi, P. H\"{a}nggi, P. Talkner, Rev. Mod. Phys.
\textbf{83}, 771 (2011).\ 

\bibitem {Jarzynski-Rev}C. Jarzynski, Annu. Rev. Condens. Matter Phys.
\textbf{2}, 329 (2011). \ \ \ \ \ \ \ \ \ \ \ \ \ \ \ \ \ \ \ \ 

\bibitem {Landau}L.D.\ Landau and E.M. Lifshitz, \textit{Statistical Physics},
Vol. 1, Third Edition, Pergamon Press, Oxford (1986).

\bibitem {Note1}To see this, we use the first law $\Delta E=\Delta_{\text{e}%
}Q+\Delta R$ and rewrite $R_{\text{diss}}$ for an isothermal work process as
follows:\ $\Delta R_{\text{diss}}=\Delta E-\Delta_{\text{e}}Q-(\Delta
E-T_{0}\Delta S)=T_{0}\Delta_{\text{i}}S$, where $\Delta_{\text{e}}Q$ is the
heat added to the system, $T_{0}$ is the temperature of the heat bath, and
$\Delta_{\text{i}}S\geq0$ is the irreversible entropy generation, a
nonnegative quantity. For a reversible process, $\Delta_{\text{i}}S=0$ so
$\Delta R_{\text{diss}}=0$. Indeed, $\Delta R_{\text{diss}}=\Delta_{\text{i}%
}W$, the accumulation of $d_{\text{i}}W=dW-d_{\text{e}}W$ in an isothermal process.

\bibitem {Note-R}By definition, $\left\langle \Delta R\right\rangle
=-\Delta_{\text{e}}W$ and differs from $-\Delta W$ due to the presence of
irreversibility $\Delta_{\text{i}}W$.

\bibitem {Jarzynski}C. Jarzynski, Phys. Rev. Lett. \textbf{78}, 2690 (1997);
C.R. Physique \textbf{8}, 495 (2007).

\bibitem {Note-SI}Any extensive or intensive quantity such as the energy $E$,
entropy $S$, volume $V$, temperature $T$, restoring force $F$, generalized
work $\Delta W$ and heat $\Delta Q$, etc. that depend on the system alone are
called system-intensive (SI) quantities. The external temperature $T_{0}$,
pressure $P_{0},$ force $F_{0}$, etc. are not SI quantities; they are
MI\ (medium intrinsic) quantities that control the exchange quantities
$d_{\text{e}}q$ for the system and determine the thermodynamic forces such as
$T-T_{0},P-P_{0},F+F_{0}$, etc. which control the system's approach to
equilibrium. The exchange work $d_{\text{e}}W=P_{0}dV$ depends on the
MI-quantity $P_{0}~$so it does not represent an SI-quantity.

\bibitem {Gujrati-GeneralizedWork}P.D. Gujrati, arXiv:1702.00455.

\bibitem {Gujrati-GeneralizedWork-Expanded}P.D. Gujrati, arXiv:1803.09725.

\bibitem {deGroot}S.R. de Groot and P. Mazur, \textit{nonequilibrium
Thermodynamics}\textbf{, }First Edition, Dover, New York (1984).

\bibitem {Prigogine}D. Kondepudi and I. Prigogine, \textit{Modern
Thermodynamics}, John Wiley and Sons, West Sussex (1998).

\bibitem {Gujrati-II}P.D. Gujrati, Phys. Rev. E \textbf{85}, 041128 (2012);
P.D. Gujrati, arXiv:1101.0438.

\bibitem {Gujrati-Entropy2}P.D. Gujrati, Entropy, \textbf{17}, 710 (2015).

\bibitem {Gujrati-Stat}P.D. Gujrati, arXiv:1206.0702.

\bibitem {Crooks}G.E. Crooks, J. Stat. Phys. \textbf{90}, 1481 (1998).

\bibitem {Seifert-PRL}U. Seifert, Phys. Rev. Lett. \textbf{95}, 040602 (2005).

\bibitem {Seifert-EPJ}U. Seifert, Eur. Phys. J. B 64, 423 (2008).

\bibitem {Jarzynski-EPJ}C. Jarzynski, Eur. Phys. J. B 64, 331 (2008).

\bibitem {Gujrati-Crooks}P.D. Gujrati, arXiv:1901:11185.

\bibitem {Strook}D.W. Strook, \textit{An} \textit{Introduction to Markov
Chains}. Springer-Verlag, Berlin (2014).
\end{thebibliography}
\end{document}